%% file: main.tex
\documentclass[conference,letterpaper,10pt]{IEEEtran}
\IEEEoverridecommandlockouts

\usepackage{cite}
\usepackage{amsmath,amssymb,amsfonts}
\usepackage{graphicx}
\usepackage{textcomp}
\usepackage{xcolor}
\usepackage{booktabs}
\usepackage{listings}
\usepackage{url}
\usepackage{tikz}
\usetikzlibrary{arrows.meta,positioning,shapes.geometric,calc,fit,backgrounds}
\usepackage{hyperref}

\hypersetup{
  colorlinks=true,
  linkcolor=black,
  citecolor=black,
  urlcolor=blue,
}

\def\BibTeX{{\rm B\kern-.05em{\sc i\kern-.025em b}\kern-.08em
    T\kern-.1667em\lower.7ex\hbox{E}\kern-.125emX}}

\definecolor{lkpick}{HTML}{1F77B4}
\definecolor{lkdrill}{HTML}{D62728}
\definecolor{lkresearch}{HTML}{FF7F0E}
\definecolor{lkgray}{HTML}{BBBBBB}

\begin{document}

\title{Agentic Method for Deterministic Validation of Legacy Code Migration}

\author{%
  \IEEEauthorblockN{Andras Ferenczi, Jordan Docherty, Mariya Bessonov,
    Matthew Findlay, and Krishna Lingamneni}
  \IEEEauthorblockA{American Express\\
    \{Andras.L.Ferenczi1, Jordan.Docherty, Mariya.Bessonov,
    Matthew.Findlay\}@aexp.com\\
    krishna.k.lingamneni@aexp.com}
}

\maketitle

\begin{abstract}
\input{sections/abstract}
\end{abstract}


\section{Introduction}
\input{sections/intro_lead}
\input{sections/intro_metaphor}
\input{sections/intro_harness_setup}
\input{sections/intro_contributions}
\input{sections/intro_scope}

\section{Related Work}
\input{sections/related_work}

\section{Method}\label{sec:how}
\input{sections/method_pipeline}
\input{sections/method_roles}
\input{sections/method_lockpicks}
\input{sections/method_drill}
\input{sections/method_recursion}
\input{sections/method_parity}
\input{sections/method_intervention}

\section{Experimental Results}
\input{sections/experiments_main}
\input{sections/experiments_production}

\section{Threats to Validity}
\input{sections/threats_intro}
\input{sections/threats_external}
\input{sections/threats_construct}
\input{sections/threats_internal}
\input{sections/threats_overfitting}
\input{sections/threats_java_oracle}

\section{Future Work}
\input{sections/future_work}

\section{Conclusion}
\input{sections/conclusion}

\bibliographystyle{IEEEtran}
\bibliography{references}

\end{document}

%% file: sections/abstract.tex
  Migration of legacy COBOL programs to Java requires extensive testing to ensure correct functionality. This effort is often complicated by the lack of test data and the difficulty of validating all corner cases. In this paper we propose a novel agentic test-synthesis method, the ``Locksmith Loop,'' which is initiated by preparing two runtime environments: the COBOL source and the generated Java target are each instrumented with mocks and executed off-mainframe on commodity hardware, then an iterative agentic loop performs Witness Search over input mocks to penetrate program branches, followed by parity-preserving mutations. When routing boundaries are reached, an analyzer identifies a Locked Paragraph: a condition preventing deeper exploration. Across three COBOL-Java case studies, spanning open-source and an internal production-like COBOL program and 430 to 4,114 source lines, Locksmith consistently improved coverage beyond input-search plateaus, reaching nearly complete coverage on the two open-source programs and 91.90\% branch coverage on the internal production-like COBOL program. The generated Java matched the COBOL reference under deterministic parity checks in all accepted test cases. Through these findings we demonstrate, to the best of our knowledge, a novel approach for validating agentic coding output using a deterministic oracle.


%% file: sections/intro_lead.tex
Many enterprises maintain large repositories of COBOL running on mainframes, carrying decades of accumulated business and proprietary processes. Agentic coding systems make it increasingly plausible to explore, reason about,
and migrate legacy systems. However, evaluating the parity between modern and legacy variants of the same system is still a difficult task. Often, both systems must be run
under strict evaluation until enough confidence is established to cut traffic over to the replacement system.

A lack of testing data is a common issue: often a limited number of real-world scenarios are covered,
leaving a long tail of corner scenarios untested. Additionally, “broken-as-usual” cases exist where the system exhibits
behavior that doesn't meet the original product definition, but has become part of it over time. Ensuring proper coverage
of these scenarios is important in lift-and-shift style modernization. In this paper we present a technique for improving
branch coverage across a COBOL software stack that has been targeted for Java migration.
The technique builds on top of established search-based test generation algorithms and introduces agentic mechanisms for
intelligent, symmetric mutation of both systems and the test harness itself.

We observed that multiple established techniques such as combinatorial interaction testing, adaptive random testing,
search-based test generation and many-objective variants, fuzzing and quality diverse algorithms such as MAP-Elites were
able to efficiently cover the input search space, but converged to similar coverage ceilings, suggesting that the remaining
branches were structurally unreachable without modification to state that lives outside of the input-search space.
Our approach identifies points where input-search is blocked and utilizes agentic mutation to push beyond this limitation
while observing parity between the legacy and target software.

%% file: sections/intro_metaphor.tex
\subsection{The Locksmith metaphor}
\label{sec:metaphor}

We use the Locksmith metaphor to help us conceptualize the distinction between branches reachable through input search and branches that can only be exposed through a parity-preserving mutation.

Each branch can be thought of as a door, and the inputs or mocked state required to cover the branch as the keys to that door. Some doors can be opened easily, while others require a drill: a parity-preserving mutation that exposes a blocked execution path. By running multiple search algorithms across reachable branches, we can identify which doors are easily reachable and which remain blocked after repeated attempts. The remaining branches may be structurally unreachable from the current harness state.

At this point, we can use a drill to expose a new region of execution. Once the new door is unlocked, we resume lockpicking, i.e.\ Witness Search, to explore the newly reachable paths. By applying this process recursively, we can systematically explore paths that couldn't be reached before, enabling us to achieve higher coverage across
the codebase, identifying edge cases that may have been missed by traditional methods.


Throughout the paper, we use the following terminology to describe the Locksmith metaphor:

\begin{itemize}
  \item \textbf{Locksmith Loop}: the end-to-end methodology.
  \item \textbf{Witness Search}: the phase that uses multiple algorithms to explore the program input space, including mocked backend responses and environmental state.
  \item \textbf{Locked Paragraph}: a COBOL paragraph that cannot be reached via Witness Search alone.
  \item \textbf{Mutation}: the code-altering phase that applies parity-preserving mutations to both the COBOL and Java targets.
  \item \textbf{Parity Gate}: the deterministic oracle, a tool that runs both targets using the same witness inputs and reports end-state discrepancies, if any.
  \item \textbf{Mutation Skills}: reusable code-mutation strategies used by the Authoring Layer to propose mutations applied to both targets. The Authoring Layer captures a previously successful mutation as an AI skill and stores it in the Skill Catalog for reuse in future iterations.
  \item \textbf{Transition (edge) coverage}: the percentage of static paragraph-to-paragraph control-flow transitions traversed at least once.
  \item \textbf{Branch coverage}: the percentage of instrumented branch outcomes that Locksmith traversed at least once.
\end{itemize}

%% file: sections/intro_harness_setup.tex
\subsection{Migration and harness setup}
\label{sec:migration-harness-setup}

\begin{figure}[t]
\centering
\resizebox{\columnwidth}{!}{%
\input{migration_harness_setup_figure.tex}%
}
\caption{Migration and harness setup. Colors denote ownership: student components are light blue; Authoring Layer components are light yellow.}
\label{fig:migration-harness-setup}
\end{figure}
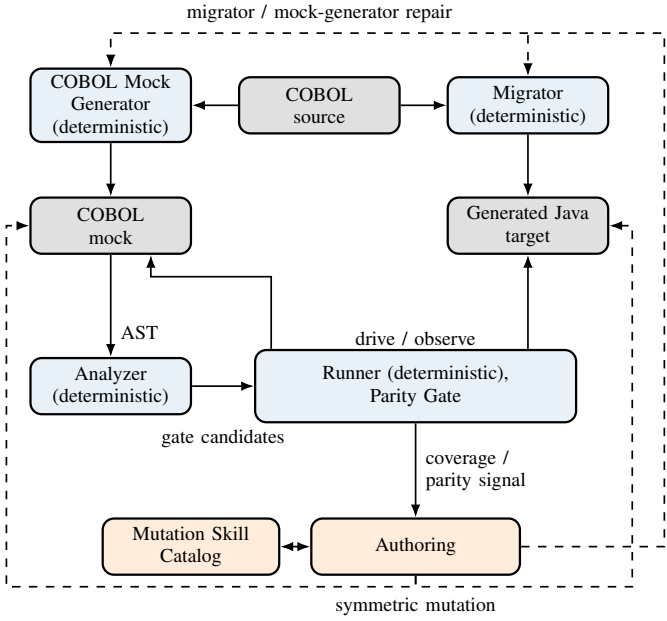

The COBOL source is passed through an internal deterministic Migrator to produce a generated Java target. A deterministic COBOL Mock Generator produces a runnable COBOL mock. The COBOL mock and generated Java target are exercised by a shared Test Harness that records a behavioral fingerprint and enforces the Parity Gate. When Witness Search stalls, the Authoring Layer uses Mutation Skills to propose symmetric mutations, meaning corresponding changes to both the COBOL mock and generated Java target.

The two actors named in Fig.~\ref{fig:migration-harness-setup} are defined next.

\label{sec:teacher-student}
The Locksmith Loop follows a teacher-student pattern. The
\emph{student} is a deterministic set of tools that carry out the migration, Witness Search,
Mutation, and parity-checking pipeline. The \emph{teacher} role is fulfilled by the
Authoring Layer, a supervisory AI agent invoked when the
student reaches a blocked execution path that ordinary input search cannot open. The
Authoring Layer does not write the migrated program directly. Instead, it selects an
existing skill or proposes a new skill to open the blocked execution path.
Correctness of each proposal is determined by the Parity Gate.
Compilation and behavioral-equivalence checks accept or reject each proposal.
A parity failure triggers the Authoring Layer to intervene: if its mutation passes the gate, the loop resumes.
This separation of generation from deterministic judgment is central to the
robustness of the loop for legacy migration. It also matches a pattern
increasingly common in agentic software engineering~\cite{bouzenia24repairagent,yang24sweagent}
and in industrial AI-assisted testing~\cite{alshahwan24testgenllm}: AI agent
proposals are rejected or accepted by deterministic execution.



%% file: migration_harness_setup_figure.tex
\begin{tikzpicture}[
  font=\scriptsize,
  box/.style={
    draw, thick, rounded corners, align=center,
    fill=lkpick!10,
    minimum width=20mm, minimum height=7mm, inner sep=2pt
  },
  artifact/.style={
    draw, thick, rounded corners, align=center,
    fill=lkgray!45,
    minimum width=20mm, minimum height=7mm, inner sep=2pt
  },
  harness/.style={
    draw, thick, rounded corners, align=center,
    fill=lkpick!10,
    minimum width=40mm, minimum height=9mm, inner sep=2pt
  },
  agent/.style={
    draw, thick, rounded corners, align=center,
    fill=lkresearch!15,
    minimum width=26mm, minimum height=7mm, inner sep=2pt
  },
  skills/.style={
    draw, thick, rounded corners, align=center,
    fill=lkresearch!15,
    minimum width=22mm, minimum height=7mm, inner sep=2pt
  },
  arrow/.style={-{Latex[length=1.6mm]}, semithick},
  botharrow/.style={{Latex[length=1.6mm]}-{Latex[length=1.6mm]}, semithick},
  dasharrow/.style={-{Latex[length=1.6mm]}, semithick, dashed}
]
\node[box] (mockgen) at (0,1.5) {COBOL Mock\\Generator\\(deterministic)};
\node[artifact] (cobolsrc) at (2.6,1.5) {COBOL\\source};
\node[box] (migrator) at (5.2,1.5) {Migrator\\(deterministic)};
\node[artifact] (cobolmock) at (0,0) {COBOL\\mock};
\node[artifact] (java) at (5.2,0) {Generated Java\\target};
\draw[arrow] (cobolsrc.west) -- (mockgen.east);
\draw[arrow] (cobolsrc.east) -- (migrator.west);
\draw[arrow] (mockgen) -- (cobolmock);
\draw[arrow] (migrator) -- (java);

\node[box] (analyzer) at (0,-2.0) {Analyzer\\(deterministic)};
\node[harness] (harness) at (3.8,-2.0) {Runner (deterministic),\\Parity Gate};
\draw[arrow] (cobolmock.south) -- node[right, yshift=-3.5mm]{\strut AST} (analyzer.north);
\draw[arrow] (analyzer.east) -- (harness.west);
\node[font=\scriptsize] at (1.4,-2.65) {gate candidates};

\draw[arrow] (harness.north -| java.south) -- (java.south);
\draw[arrow] ([xshift=-18mm]harness.north) -- ++(0,0.9)
  -| ([xshift=5mm]cobolmock.south);
\node[font=\scriptsize] at (3.8,-1.4) {drive / observe};

\node[agent] (teacher) at (3.8,-4.0) {Authoring};
\node[skills] (skills) at (1.0,-4.0) {Mutation Skill\\Catalog};
\draw[arrow] (harness.south) -- (teacher.north);
\node[font=\scriptsize, align=left] at (4.55,-3.05) {coverage /\\parity signal};
\draw[botharrow] (skills.east) -- (teacher.west);

\draw[dasharrow] (teacher.south) -- (3.8,-4.5) -- (-1.3,-4.5)
  -- (-1.3,0) -- (cobolmock.west);
\draw[dasharrow] (teacher.south) -- (3.8,-4.5) -- (6.5,-4.5)
  -- (6.5,0) -- (java.east);
\node[font=\scriptsize] at (3.8,-4.75) {symmetric mutation};

\draw[dasharrow] (teacher.east) -- (6.9,-4.0) -- (6.9,2.4) -- (5.2,2.4) -- (migrator.north);
\draw[dasharrow] (5.2,2.4) -- (0,2.4) -- (mockgen.north);
\node[font=\scriptsize] at (2.6,2.65) {migrator / mock-generator repair};
\end{tikzpicture}

%% file: sections/intro_contributions.tex
\subsection{Contributions}
\label{sec:contributions}

This paper makes three contributions:

\begin{enumerate}
  \item Proposes an effective solution to improve quality of migrated code by introducing a novel testing methodology that is performed as an addition to the existing Software Development Lifecycle (SDLC) and which significantly reduces the scenarios that integration and quality assurance (QA) testing cannot detect. While our examples are restricted to COBOL-to-Java migrations, these principles can
  be applied to other use cases, such as Java-to-Kotlin or C\#-to-Java migrations, where a strict 1:1 behavioral equivalence between source and target is feasible. The approach further extends to cross-paradigm migrations, e.g., C-to-Go, C-to-Java, or COBOL-to-Java as demonstrated here, provided the migrated code preserves the original input-output behavior. Each such scenario must be evaluated individually to determine the extent of behavioral coverage achievable.

  \item Provides an early concrete example of the shifting role engineers will have in the new era of agentic software development, away from low-level code design and implementation details toward defining outcome goals and deterministic validation criteria for migration outcomes. The parity-as-oracle design provides an explicit acceptance mechanism for checking behavioral equivalence between the existing legacy system and the migrated code.

  \item Builds on agentic software-engineering and test-improvement patterns~\cite{bouzenia24repairagent,alshahwan24testgenllm} and uses closed-loop refinement to capitalize on the AI agent's ability to improve through repeated execution feedback. The experiments we conducted ran autonomously while iteratively improving after each cycle. The Authoring Layer adjusted the migrator as it encountered new scenarios to improve the generated Java target, learned from the mutations, and captured new scenarios into reusable AI skills. In our experiments, the loop continued unattended for hours to produce the presented results.
\end{enumerate}


%% file: sections/intro_scope.tex
\subsection{Scope and assumptions}
\label{sec:scope}

The technique is evaluated in a specific migration scenario: a 1:1 migration of COBOL systems to Java. The primary success criterion for this paper is increasing branch coverage
with validation of parity under the same inputs, rather than system overhaul. Our migration harness has
opinionated infrastructure substitutions for replicating legacy dependencies, e.g. legacy RDBMS instances are replaced with PostgreSQL, MQ instances are replaced with
RabbitMQ. The strategy we outline is twofold: first port the application with parity, then refactor once confidence is established in the
new system.


%% file: sections/related_work.tex
\label{sec:related}

Our proposed Locksmith Loop lies in the intersection of several lines of research.
We will focus the discussion on four key technical areas and indicate our contributions
in each of these settings.

\paragraph*{1. Test-suite amplification and harness testability.}
Test-amplification literature follows the standard view where
tests are seen as artifacts and iteratively improved upon rather
than generated a single time~\cite{danglot17amplification,danglot18dspot}.
In DSpot~\cite{danglot18dspot}, this iterative view is concretely realized,
developer-written tests are augmented and fed back as
merge-ready patches.
EvoSuite and EvoSuiteAmp are closely related systems, in which existing
tests are leveraged as seeds for search-based
improvement~\cite{fraser11evosuite,roslan22evosuiteamp}.
While the Locksmith Loop maintains this iterative stance,
it updates the transformation target by expanding the
harness interface itself,
rather than keeping it fixed. It updates the inputs that can
be expressed by the harness and re-runs Witness Search from within
this newly expanded input space.
This technique places the Locksmith adjacent to work in software testability
and testability transformation~\cite{harman04testability,garousi18testability}.
However, the Locksmith expands the harness boundary in lieu of rewriting
the program subject to testing.
The Locksmith iterates upon the controllability boundary itself.

\paragraph*{2. Search-based input generation, fuzzing, and symbolic execution.}
Witness Search builds on established techniques in automated input-space exploration,
as surveyed in~\cite{manes18fuzzing}, including combinatorial-
interaction testing~\cite{kuhn04combinatorial}, adaptive random
testing~\cite{chen10art}, many-objective search-based testing
(MIO~\cite{arcuri19mio}), quality-diversity exploration via
MAP-Elites~\cite{mouret15mapelites}, and its bandit-aware refinement
Monte Carlo Elites~\cite{sfikas21mcelites}. We run these algorithms in parallel.
Empirical studies of continuous fuzzing, such as OSS-Fuzz~\cite{ding21ossfuzz}
demonstrate nontrivial progression dynamics, including bugs found in spiky patterns of
slow growth followed by rapid bursts.
Symbolic and concolic execution suggest complementary strategies for input-space
exploration from the angle of path conditions and constraints~\cite{baldoni16symexec}.
Hybrid systems like Driller invoke selective concolic execution when fuzzing gets stuck~\cite{stephens16driller}.
Program-transformation fuzzers like T-Fuzz~\cite{peng18tfuzz} employ code mutation to break past hard input checks for the
purpose of bug discovery rather than parity-gated migration validation.
The Locksmith Loop initially follows the input generation and fuzzing patterns and later moves on from it
once saturated by identifying a plateau once multiple algorithms converge to within $\pm$2--3 branches of the same
count. Rather than augmenting the budget, the next step is a \emph{mutation of the harness boundary} to continue the search from an updated angle.

\paragraph*{3. Differential and metamorphic test oracles.}
Classic differential testing, recently enhanced with LLM assistance such
as in Mokav~\cite{etemadi24mokav} and DiffSpec~\cite{rao24diffspec},
takes two code implementations for the same spec and flags discrepancies in outputs as bugs.
When the exact outputs are unavailable, metamorphic testing~\cite{chen18metamorphic}
checks for relations between inputs and outputs across executions.
Locksmith Loop's Parity Gate judges candidate inputs, and it also acts as a guardrail for
mutating the harness, keeping the mutations that both preserve parity and expand the reachable input space.

\paragraph*{4. Program repair and agentic LLM software engineering.}
Locksmith follows the propose-validate pattern as in automated program repair (APR)
and agentic software engineering~\cite{zhang23apr,yang25llmapr,jimenez23swebench,yang24sweagent}.
RepairAgent~\cite{bouzenia24repairagent} invokes LLM tool usage
and validates fixes against program feedback;
TestGen-LLM~\cite{alshahwan24testgenllm} deploys
LLM-generated tests only if deterministic filters confirm they build, pass, and improve coverage;
TestPilot~\cite{schafer23testpilot} uses LLM to generate tests and re-prompting if generated tests fail;
Mut4All~\cite{wang25mut4all} uses LLM agents to synthesize compiler-fuzzing mutators from bug reports.
Locksmith applies a similar structure to \emph{harness mutation}, where the LLM proposes the edits but
the deterministic analyzer / runner / parity oracle decide which edits survive.

\paragraph*{Legacy and mainframe modernization}
LLM-based mainframe modernization is an active area of research, including
XMainframe~\cite{dau24xmainframe}, COBOL-Coder's domain-adapted COBOL
generation and translation benchmarks~\cite{dau26cobolcoder},
COBOL-to-Java refinement~\cite{gandhi24coboljava}, and enterprise-scale
COBOL-to-Java pipelines that combine program analysis with LLMs~\cite{chakravarthy26enterprise}.
A close comparison for validating COBOL-to-Java transformation uses symbolic
execution to generate unit tests for COBOL, translates them into JUnit tests
with mocking, and checks for semantic
equivalence~\cite{hans25cobol2javatesting,kumar25cobol2java}.
While targeting this same equivalence problem, Locksmith treats validation as an
internal gate in a search-and-mutation loop rather than a final post-hoc check.

\paragraph*{Position}

To summarize, Locksmith is a novel combination of
these four key technical ideas that are typically studied in isolation:
test-suite amplification, search-based input generation, differential testing, and agentic software engineering.
At the heart of Locksmith is the recursive Witness Search and Mutation Loop that expands
the reachable input space until it saturates, then mutates the harness to
open new regions for exploration. We are not aware of prior COBOL migration work addressing legacy-to-modern migration
using recursive harness expansion as the central validation mechanism.


%% file: sections/method_pipeline.tex
\subsection{Pipeline overview}
\label{sec:pipeline}

The pipeline begins with source-level transformations that produce a runnable
\emph{mock COBOL} program. In this program, external operations such as file
I/O, \texttt{CALL}, \texttt{EXEC SQL}, and \texttt{EXEC CICS} are mocked and
backed by either COBOL index files or a relational database such as
PostgreSQL. The compiled mock binary supplies the branch-coverage measurements
used throughout the loop.

Next, the legacy program's Abstract Syntax Tree (AST) is produced by an external parser. 
The AST is then provided as input to a deterministic Migrator.
 Its Java code generator emits one
class per section and one method per paragraph, and maps COBOL control
flow (\texttt{PERFORM}, \texttt{GO TO}, \texttt{ALTER}, and
\texttt{EVALUATE}) to equivalent Java control flow. The COBOL mock and
the generated Java target are run against the same test cases. The outputs
of these test cases are compared as seen in Section~\ref{sec:parity}.

Each test case consists of an \emph{input-state} and \emph{stub-state}. 
The input-state specifies the variable assignments applied during execution, 
while the stub-state supplies the records returned by mocked external operations. 
After test execution, the system records a \emph{behavioral fingerprint} containing 
the paragraphs entered, branches taken, stub-operation log, and terminal observable state. 
This fingerprint is used for coverage tracking and parity comparison.

The pipeline then loops through two phases. The Witness Search
phase applies Witness Search until it reaches a coverage
plateau. The Mutation phase applies a candidate
mutation to reach a locked region, then keeps the mutation only if it
adds coverage and passes the Parity Gate. After a successful code mutation,
the loop returns to Witness Search from the newly reached region.
Figure~\ref{fig:loop} gives the recursive control structure.


%% file: sections/method_roles.tex
\subsection{Four components: Migrator, Analyzer, Runner, and Authoring}
\label{sec:teacher-components}

Within the teacher-student pattern, the student comprises the
deterministic execution components, while the teacher role is fulfilled by the
Authoring Layer. Figure~\ref{fig:migration-harness-setup} gives the architecture-level mapping.

\paragraph{Deterministic Migrator}
The migrator is a deterministic component that translates the COBOL source into the Java target.

\paragraph{Deterministic Analyzer}
The analyzer is a deterministic component that examines the AST and the live
mock, then identifies gates from program
structure.

\paragraph{Deterministic Runner}
The runner compiles the binary, executes test cases, measures
branch coverage, and applies the Parity Gate along the three
equivalence axes in Section~\ref{sec:parity}. A code mutation is retained
only when it increases branch coverage.

\paragraph{Agentic Authoring}
The Authoring Layer is invoked when the deterministic analyzer reaches a
blocked execution path. It then synthesizes a new skill that enables
the system to proceed beyond the blockage. When the Parity Gate reports a
divergence, the Authoring Layer may also patch the migrator's Java code generator.

\begin{figure}[t]
\centering
\begin{tikzpicture}[
  font=\scriptsize,
  node distance=4.6mm and 7mm,
  >=Latex,
  pick/.style={draw=lkpick!75, line width=0.7pt, rounded corners,
               fill=lkpick!10, align=center, inner sep=3.2pt,
               text width=52mm},
  drill/.style={draw=lkresearch!85, line width=0.7pt, rounded corners,
                fill=lkresearch!15, align=center, inner sep=3.2pt,
                text width=52mm},
  gate/.style={draw=lkpick!75, line width=0.7pt, rounded corners,
               fill=lkpick!10, align=center, inner sep=3.2pt,
               text width=52mm},
  dec/.style={draw=gray!70, line width=0.6pt, diamond, aspect=2.4,
              fill=gray!8, align=center, inner sep=0.5pt,
              text width=26mm},
  term/.style={draw=gray!75, line width=0.7pt, rounded corners,
               fill=gray!14, align=center, inner sep=3.2pt},
  lbl/.style={font=\scriptsize\itshape, inner sep=1.5pt},
]
\node[term] (start) {Bare mock CBL \;$+$\; generated Java target};
\node[pick, below=of start] (bfs)
  {\textbf{Witness Search}\\[1pt]
   six witness algorithms try every reachable paragraph\\
   pairwise $\cdot$ 3-way $\cdot$ LHS $\cdot$ ART $\cdot$ MAP-Elites $\cdot$ UCB1};
\node[dec, below=of bfs] (plateau) {new\\ branches?};
\node[pick, below=6.5mm of plateau] (rank)
  {rank remaining Locked Paragraphs by uncovered-branch count};
\node[dec, below=of rank] (any) {Locked Paragraph\\ left?};
\node[drill, below=6.5mm of any] (drill)
  {\textbf{code mutation}\\[1pt]
   Authoring Layer proposes a Mutation Skill;
   student applies to \emph{both}
   COBOL and Java};
\node[pick, below=of drill] (focus)
  {Post-Mutation Witness Search (UCB1) on the mutated binary};
\node[dec, below=of focus] (gain) {coverage\\ gain?};
\node[gate, below=6.5mm of gain] (parity)
  {\textbf{Parity Gate}\\[1pt]
   run latest test case through the generated Java target;\\
   COBOL\,$\equiv$\,Java $\Rightarrow$ keep \quad else log divergence};
\node[term, below=4.6mm of parity] (done)
  {\textbf{done}: every Locked Paragraph attempted};

\draw[->] (start) -- (bfs);
\draw[->] (bfs) -- (plateau);
\draw[->] (plateau) -- node[lbl,right]{no (plateau)} (rank);
\draw[->] (rank) -- (any);
\draw[->] (any) -- node[lbl,right]{yes} (drill);
\draw[->] (drill) -- (focus);
\draw[->] (focus) -- (gain);
\draw[->] (gain) -- node[lbl,right]{yes} (parity);
\draw[->] (parity) -- (done);

\draw[->] (plateau.east) -- ++(18mm,0)
  node[lbl,above,pos=0.55]{yes} |- (bfs.east);
\draw[->] (any.east) -- ++(16mm,0)
  node[lbl,above,pos=0.55]{no} |- (done.east);
\draw[->] (gain.east) -- ++(18mm,0)
  node[lbl,above,align=center,pos=0.5]{no: revert,\\ mark failed} |- (rank.east);
\draw[->,line width=1pt] (parity.west) -- ++(-20mm,0)
  node[lbl,align=center,pos=0.5,above,xshift=-5mm]{recurse:}
  node[lbl,align=center,pos=0.5,below,xshift=-5mm]{search from inside} |- (bfs.west);
\end{tikzpicture}
\caption{The Locksmith Loop. The \textbf{Witness Search} phase
sweeps reachable paragraphs with six Witness Search algorithms, including UCB1~\cite{auer02ucb1}, until
two consecutive rounds add no branches. On plateau, the \textbf{code mutation}
phase opens one Locked Paragraph through a parity-preserving
mutation proposed by the Authoring Layer and applied by the student to both the COBOL and its generated Java target. Post-Mutation
Witness Search confirms the coverage gain, and the \textbf{Parity Gate}
checks agreement. A successful code mutation returns control to a full
sweep from the newly opened region. A code mutation that yields no
gain is reverted and its Locked Paragraph is marked failed. The loop terminates when all identified Locked Paragraphs have been attempted. Colors denote ownership: student
components are light blue, the Authoring Layer component is light yellow, and
control-flow nodes are gray.}
\label{fig:loop}
\end{figure}
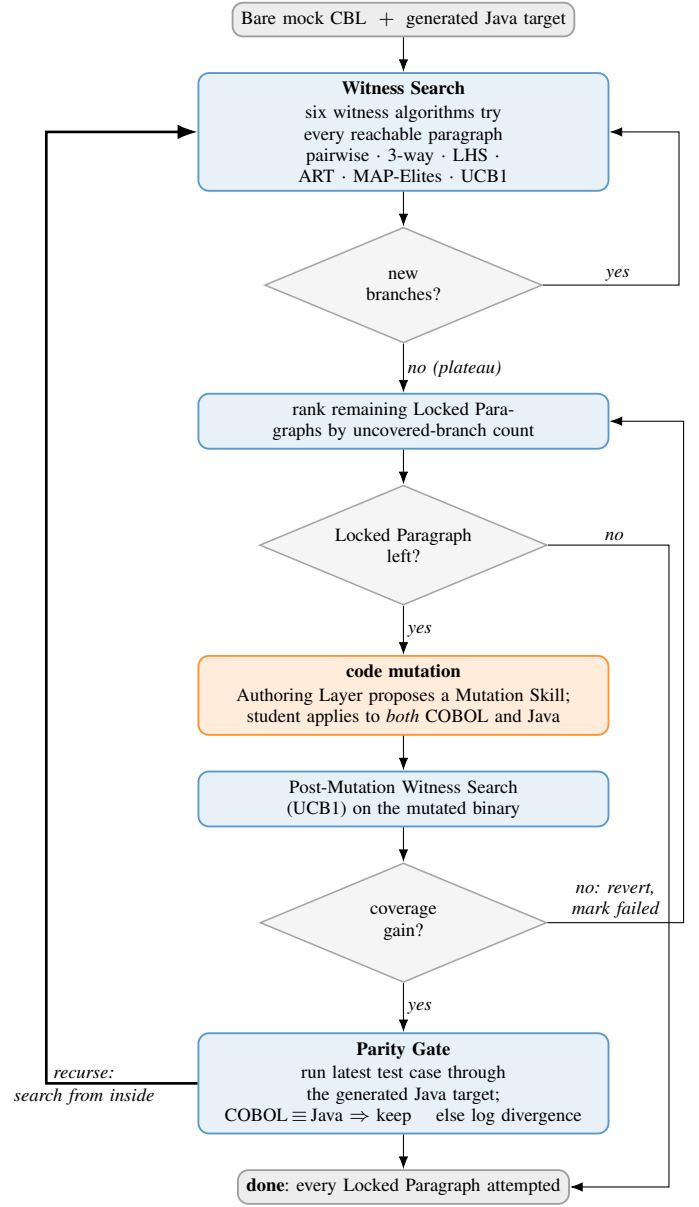
%


%% file: sections/method_lockpicks.tex
\subsection{Witness Search: input-space Witness Search}
\label{sec:lockpicks}

The Witness Search sweep runs six independent Witness Search algorithms
against the current-iteration COBOL mock binary, each run starting
from the same baseline coverage.
Each searches for execution scenarios that will open new branches. A scenario consists of
input records, initial working-storage values, and mocked external \texttt{CALL}
file-status values and return codes.
The harness catalog tracks allowable choices for each scenario, and these six algorithms only differ
in how they explore the combinatorial space of these choices. They are:

\begin{itemize}
\item \textbf{Pairwise interaction testing} ($m_{11}$): given $n$
scenario components with $|D_i|$ value choices for each field,
generate a small test
set covering every possible pair $(v_i \in D_i,\, v_j \in D_j)$ for $i < j$
(optimal size $O(|D|^2 \log n)$). Since many COBOL branches are gated by compound \texttt{IF}s or
\texttt{EVALUATE WHEN} clauses on two correlated fields (status
code $\times$ record type, account kind $\times$ balance sign), such a
pairwise-covering set surfaces each
two-field combination at least once.
\item \textbf{Three-way interaction} ($m_{16}$): pairwise interaction extends
to triples, uncovering the
smaller class of branches requiring three coordinated
values. For example, tail-handling paragraphs gated on
status $\times$ record type $\times$ end-of-file flag are covered by three-way
but not necessarily by pairwise.
\item \textbf{Latin hypercube sampling} ($m_{17}$): suitable for quasi-continuous
scenario components
 (e.g. record counts, file lengths, monetary amount), here we partition each
domain into $N$ bins and sample one value per bin,
with bins aligned across components to maximise distance-based
diversity. LHS spreads samples across the full range rather than
clustering them in a particular region.
\item \textbf{Adaptive random testing} ($m_{18}$): pick the next
input that maximises the minimum distance to previously
executed inputs, biased toward unexplored regions of the input
space. This is particularly useful when an unfamiliar program offers no
obvious structure to exploit.

\item \textbf{MAP-Elites} ($m_{19}$): a quality-diversity
algorithm~\cite{mouret15mapelites} that maintains a grid of
behavioral-descriptor cells containing the highest-fitness input
seen to map to that cell, where fitness is number of newly covered branches. COBOL
coverage profiles may cluster into a small number of
execution \emph{shapes} (e.g. clean end-of-file path, mid-stream
truncation, status-error early exit, multi-record-type batch, etc.);
MAP-Elites retains one good test case per shape, helping to expose rare branches.

\item \textbf{Upper-confidence-bound bandit} ($m_{20}$, UCB1):
treats each value choice as a bandit arm. At each step,
select the arm $a$ maximising
$\bar{x}_a + c\sqrt{(\ln t) / n_a}$, where $\bar{x}_a$ is the
empirical mean reward defined by fresh branches, $n_a$ is number of times $a$ has been selected,
and $c$ is an exploration constant. In contrast to interaction-based methods above,
 UCB1 has no combinatorial fan-out. Rather, it learns during a run which values
 tend to expose new branches. In COBOL mocks, this favors informative scenario
 components which gate many branches, such as
 status codes and record types.

\end{itemize}

Each algorithm produces a test-case set and a coverage set.
The output of one Witness Search sweep consists of the union of branches
discovered by the six algorithms, together with the test-case set.
 The next sweep will start from the best set, continuing until two consecutive
 sweeps yield no new branches, which we refer to as the
\emph{Witness Search plateau}. Once a mutation expands the harness, Locksmith
runs the six-algorithm sweep again, as the newly reachable regions may favor different search biases.
These six algorithms were selected through empirical experimentation and are not guaranteed to be optimal for all migrations; they should be treated as a representative set rather than a fixed prescription.


%% file: sections/method_drill.tex
\subsection{Mutation: parity-preserving harness mutation via skills}
\label{sec:drill-defn}

We start Witness Search with the goal of increasing code coverage by generating new inputs that facilitate discovery of new execution paths. Eventually, Witness Search reaches a routing boundary that it cannot cross (the Witness Search plateau). Next, the Mutation phase closes this gap by altering the code in order to eliminate these barriers. Unlike classical mutation testing, which injects bugs into the application to ensure proper error handling \cite{wang24llmmutation}, the Locksmith method discovers and force-opens new execution paths to ensure parity between the legacy (COBOL) and new (Java) functionality; both codebases are modified in tandem with the goal of matching their call traces as well as their outputs. In the context of the Mutation phase, the deterministic analyzer identifies candidate gates, and the Authoring Layer generates Mutation Skills that satisfy them, proposing auditable ways to reach the target paragraph. A skill, for example, may add a new dispatcher route, expose an external value through a side channel (a harness-level stub hook that supplies values the program would normally read from external sources), or force execution of a paragraph that is otherwise difficult to reach. We persist skills between rounds instead of direct code modifications in order to maintain a clear separation between the original program logic and the coverage-enhancing changes. Following the Mutation phase, previously accepted test cases are re-executed, and the mutations are kept for further mutation iterations. Because a single paragraph may be blocked by multiple independent conditions, the framework evaluates each identified gate in turn and retains only those mutations that produce additional coverage. This allows the system to systematically expand coverage while continuously verifying cross-language behavioral equivalence.
 
Mutation Skills can have hierarchical dependencies, and the resulting mutations are applied in the required order. In practice, we have found two skill types sufficient for most scenarios. The first, a dispatcher-arm skill, extends the harness so that values normally read from external sources can be supplied directly through the side channel. This is particularly effective for status-field-driven logic where execution depends on values that are otherwise difficult to control. The second, a call-injection skill, forces execution of a target paragraph from a known point in the main program flow. This helps exercise logic that is normally reached only during cleanup processing, end-of-file handling, or other uncommon execution paths. Regardless of the skill type, the same change is applied to both the COBOL and Java implementations, and parity is verified by comparing the execution path taken, external operations performed, and final observable state (Section~\ref{sec:parity}).
 
The deterministic analyzer of Section~\ref{sec:teacher-components} is responsible for deciding how to approach a coverage gap; in effect, it is a static Abstract Syntax Tree (AST) reader. For each target paragraph, it examines the control-flow conditions that prevent the paragraph from being executed and identifies the gates that need to be satisfied to reach it. When a gate depends on a specific external value, the analyzer determines the required value and provides it to the side channel so it can be pinned during execution. The analyzer always prefers solutions that work through existing program flow, such as dispatcher-based mutations, because they allow the paragraph to be reached naturally. Only when no suitable dispatcher-based approach is available does it recommend a call-injection skill. This ensures that every paragraph represented in the AST has at least one candidate mutation that can be evaluated. The analyzer is intentionally conservative and only recommends mutations that can be clearly explained in terms of a controllable variable and a known set of values. This keeps every recommendation understandable, auditable, and easy to validate before any mutation is applied.
 
\paragraph{Sibling gates within one Locked Paragraph}
A target paragraph may not always be blocked by a single condition. In many cases, multiple control-flow decisions must be satisfied before the paragraph is reached, such as a nested IF condition, an enclosing PERFORM boundary, or a specific EVALUATE branch. To maximize coverage, the analyzer identifies all such gates and generates a candidate mutation for each of them. Rather than stopping after the first successful mutation, the framework evaluates every identified gate because each one can potentially unlock additional execution paths. When a mutation results in new coverage, it becomes part of the active baseline and subsequent mutations continue from that expanded state. If a mutation does not provide any additional coverage, it is reverted. A paragraph is considered successfully opened if at least one mutation improves coverage.


%% file: sections/method_recursion.tex
\subsection{The recursion}
\label{sec:recursion}
%

The Locksmith Loop alternates between
(1) a Witness Search sweep and (2) a Mutation step.
The Witness Search sweep searches through the reachable input space,
and when it reaches a plateau, the Mutation step takes over and
mutates the harness. The process repeats,
until all identified Locked Paragraphs have been attempted.

A mutation must satisfy the following conditions to be kept: (1)
it must add coverage, and (2) it must pass the Parity Gate.
If both are met, Locksmith keeps the harness mutation
and proceeds to another Witness Search sweep, which may expose additional
Locked Paragraphs. Those new Locked Paragraphs are ranked and mutated in the same way.
If either condition fails, Locksmith reverts the mutation and
records the attempt. Locksmith does not re-attempt a Locked Paragraph that has been
attempted, thus guaranteeing that the recursion will eventually terminate.
In this setup, each recursive step either retains a parity-preserving
mutation that uncovers new branches or permanently removes
an attempted Locked Paragraph from further consideration. When all Locked Paragraphs
have been attempted, Locksmith runs a final Witness Search sweep to recover
any final branches.

The default paragraph-ranking policy is \emph{bang $\times$ feasibility},
in which a Locked Paragraph is scored as the sum of its uncovered branch count
with the proportion of its condition variables that are already
routable through the harness. Thus Locked Paragraphs that are valuable and
actionable are favored. In our experiments, we use the
\emph{greedy} policy, which ranks Locked Paragraphs only by uncovered branch count,
prioritizing deeper paragraphs with higher payoff.

The \textsc{SkillFor} step is treated as a lookup, but in practice the
catalog may be incomplete. If a selected Locked Paragraph does not already have a
corresponding skill, the Authoring Layer either develops a new skill from its
gate analysis or generates a structured \texttt{needs\_new\_skill}
record containing
the target paragraph, uncovered branches, branch conditions, and active
call stack.


%% file: sections/method_parity.tex
\subsection{The Parity Gate}
\label{sec:parity}

Every test case the COBOL side accepts (input-state $+$ stub-outcomes)
is driven through the generated Java target, and the resulting
behavioral fingerprints are compared. The gate is a differential
oracle in the sense of recent LLM-assisted differential-testing
work~\cite{etemadi24mokav,rao24diffspec}, and its acceptance of a
\emph{compatibility} relation in lieu of a ground-truth specification
fits the broader oracle-engineering position taken by the
metamorphic-testing review~\cite{chen18metamorphic}. Because each
mutation is realized on both sides, parity is checked
\emph{continuously}, after every break, not as a terminal round. Three equivalence axes are declared per-skill
in the parity contract:

\begin{itemize}
\item \texttt{paragraphs\_hit}: set comparison of paragraphs entered;
\item \texttt{stub\_log}: ordered comparison of the consumed
external-operation sequence, with whitespace-normalised value matching;
\item \texttt{terminal\_state}: pointwise comparison of observable
variable values after the run, ignoring runtime-private bookkeeping.
\end{itemize}

Divergences are emitted as structured records on the parity
intervention channel, carrying the divergence kind, a human-readable
description, and machine-readable diff data; these records form the
input contract for the optional repair step. The loop records and
proceeds; it does not block on a divergence unless configured to do
so. In our experiments the gate holds on every accepted test case
across all mutation tiers (Section~\ref{sec:experiments}).


%% file: sections/method_intervention.tex
\subsection{Authoring Layer intervention points}
\label{sec:intervention}

When the analyzer or runner reaches a decision point that it cannot
resolve, it writes a structured record at a defined intervention
point and continues. Records can therefore be consumed asynchronously by a
human operator or AI agent.


%% file: sections/experiments_main.tex
\begin{figure*}[ht]
\centering
\includegraphics[width=0.82\linewidth]{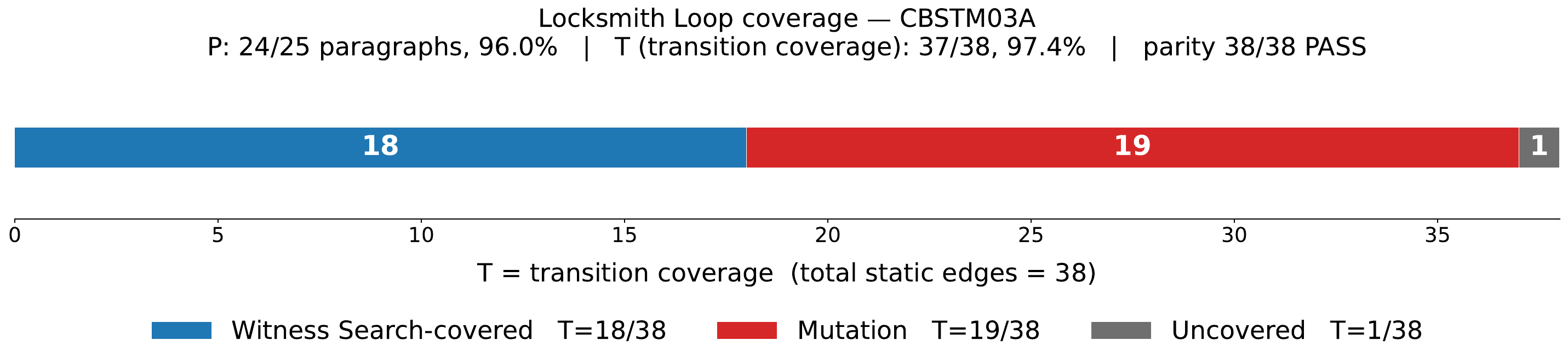}
\caption{Coverage progression for the Locksmith Loop on the test
program. }
\label{fig:progression}
\end{figure*}

\label{sec:experiments}
We experimented with the Locksmith Loop on three COBOL codebases. First, the
smaller open-source AWS CardDemo program CBACT01C (430 source lines, 16
paragraphs, 28 transitions, 62 branch probes); second, the medium-sized CardDemo
program CBSTM03A~\cite{carddemo}, publicly available, having 924 source lines,
25 paragraphs, 38 paragraph-edge transitions, and 146 branch probes; and third,
a real production-like batch program (Section~\ref{sec:prod-run}) having
4{,}114 source lines, 142 paragraphs, 146 paragraph-edge transitions, and 432
statically counted branches. The production-grade program was evaluated without
customer data, using only harness-generated test inputs and mocked external
responses. The progression charts for CBSTM03A and the
production-grade program share the same metric definitions: P is paragraphs hit
over total paragraphs in the AST; T is observed-over-total static paragraph
edges. Each edge is classified by color: EXIT/GOBACK edges follow the Directed
Acyclic Graph (DAG) color of their destination node (range-completion rule: an
EXIT terminator counts as covered once its source paragraph runs), and non-EXIT
edges follow the phase that first observed the transition (blue: Witness Search,
red: Mutation; red takes precedence on overlap). The \emph{parity $X$/$N$ PASS}
annotation in each chart reports the count of accepted test cases for which the
generated Java target reproduces the COBOL run's paragraphs hit, external
effects, and outputs exactly, i.e.\ the fraction of test cases (TCs) that survive
the Parity Gate. A \emph{force-set mutation} is a parity-preserving mutation
that, at entry to a chosen paragraph, overrides the value that would have been
mock-returned from an external operation (file read, database fetch,
message-queue receive, subprogram call) with a chosen value, applied
symmetrically to the COBOL mock and its generated Java target. We also documented
the threats to validity in Section~\ref{sec:threats}. In this experiment we
present the contributions of each method as follows: Figure~\ref{fig:progression}
shows the progression by phase and Section~\ref{sec:unreached} discusses the
residual uncovered branches and their structural causes.

\begin{figure}[t]
\centering
\includegraphics[width=\linewidth]{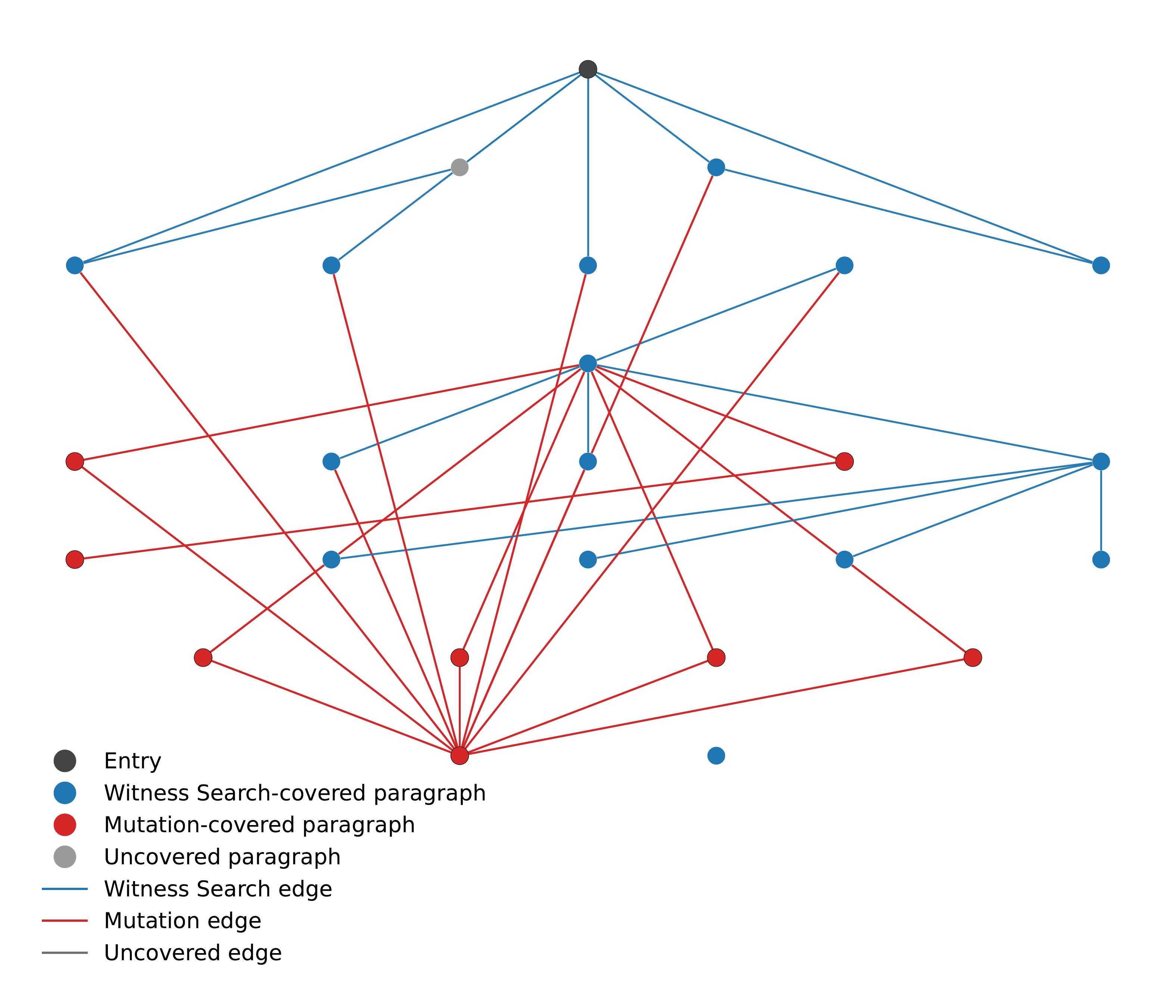}
\caption{Control-flow graph of the test program~\cite{carddemo}. Nodes are
paragraphs; edges are static paragraph-to-paragraph control-flow transitions.
Node color marks the phase that first reached the paragraph; isolated blue/gray
nodes are expected because sequential COBOL fall-through is not modeled as an
explicit edge. Edge colors: \textbf{Blue}: observed by Witness Search,
\textbf{Red}: observed after a mutation (red takes precedence on overlap),
\textbf{Gray}: not reached by this run.}
\label{fig:dag}
\end{figure}

All of the six Witness Search algorithms (Section~\ref{sec:lockpicks}) end
roughly with the same coverage ($\pm$2--3 branches). We generalize, based on
empirical observation, that this boundary is a structural one rather than
algorithmic (we experimented with even more algorithms that were discarded due
to poor performance). Based on our observation, Witness Search provides most of
the benefits but Mutation manages to penetrate into paragraphs that are only
reachable via code mutations due to program-specific conditionals that cannot be
reached otherwise. In the first example we managed to reach previously Locked
Paragraphs this way: the Mutation phase opened 8 paragraphs and 19 transitions
that Witness Search alone could not reach, which are exactly the 19 red
transitions shown in Figure~\ref{fig:dag}, which colors each transition by the
first phase to traverse it., after which a post-mutation sweep explores the newly
opened region. Mutation's value lies as much in the region it exposes as in the
branches it directly unlocks. At depth-2, after opening the depth-1 gates and
proceeding with Witness Search, the agent opens the gates on the second layer
and this triggers the six algorithms to find a new set of witnesses to penetrate
this layer too. Defects detected during each pass of the loop are fixed in the
migrator rather than in the produced code through Authoring Layer-proposed
migrator fixes, thus achieving parity. In the end, all 38 test cases passed the
Parity Gate across the three equivalence axes: paragraphs entered,
external-operation sequence, and terminal state.

Of the 146 branch probes in the final instrumented mock for CBSTM03A, 38 remained
uncovered, corresponding to 74.0\% branch coverage. Note that
Figure~\ref{fig:dag} reports paragraph and transition coverage, not branch
coverage. With sufficient patience and significant human intervention we could
have achieved better results, and even at the current coverage we are
exceeding any alternative method's output that we are aware of. At this point we
are not aware of structural reasons preventing 100\% coverage aside from the
complexity of the mocking. The existing coverage was achieved via experimentation
using frontier models available at the time of this writing. For example, the
Authoring Layer (coding agent) failed to apply learnings from one skill to open
the gates for a related scenario, and human intervention was required to identify
the issue. The expectation is that, as new models are released, autonomous
coverage runs will yield incrementally better results over time.
\label{sec:unreached}

Figure~\ref{fig:dag} shows the open-source~\cite{carddemo} program's static
control-flow graph, with each paragraph colored by the phase that first reached
it. Witness Search (input-space exploration) is the first step, shown in blue.
The red portion is the result of Mutation (code mutation); the gray region is
the uncovered portion of the program. A Witness Search sweep follows every successful
mutation to discover newly-exposed branches.

On the smaller CBACT01C program (430 lines, 62 branch probes), the loop achieved
$100\%$ paragraph coverage (16/16), $100\%$ transition coverage (28/28), and
$96.8\%$ branch coverage (60/62), confirming that near-complete saturation is
attainable on programs of moderate complexity.

%% file: sections/experiments_production.tex
\subsection{A production-shape run}
\label{sec:prod-run}
Our second experiment was performed on a significantly larger production-grade COBOL program having 4{,}114 source lines, 142 paragraphs, and 432 statically counted branches. The goal was to validate that the Locksmith Loop scales to much larger and more complex programs as well. Our experiment produced a paragraph coverage of $\mathrm{P}{=}135/142$ ($95.1\%$) and a transition coverage of $\mathrm{T}{=}101/146$ ($69.2\%$). Branch coverage reached $91.90\%$ ($397/432$).

The loop progressed as follows. The Witness Search phase reached $\mathrm{P}{=}110/142$ paragraphs and $\mathrm{T}{=}54/146$ connections; the Mutation phase added $\Delta\mathrm{P}{=}{+25}$ paragraphs and $\Delta\mathrm{T}{=}{+47}$ connections before plateauing at $\mathrm{P}{=}135/142$, $\mathrm{T}{=}101/146$. Witness Search contributed 61 raw executions and Mutation contributed 105, for 166 combined executions. Throughout the experiment, after each TC a parity check was performed and when discrepancies were detected a new set of fixes were proposed by the Authoring Layer and applied to the migrator and/or the COBOL Mock Generator, if applicable. The types of discrepancies we encountered include: end-of-stream and termination semantics, control-flow fidelity, program-state and data-layout fidelity, data typing and value normalization, database and mock-backend behavioral equivalence, file I/O side-effect and ordering equivalence, fixture and environment consistency, measurement and normalization artifact control. In the end, the generated Java target reproduced the COBOL behavior on every parity axis (set of paragraphs hit, same external effects, same outputs
produced): \textbf{166/166 PASS, 0 FAIL, 0 ERROR}.
Notably, the loop progressed autonomously without any human intervention. Agentic coding software needs deterministic validation in order to produce reliable outputs and we achieved exactly that.

\begin{figure*}[t]
\centering
\includegraphics[width=0.82\linewidth]{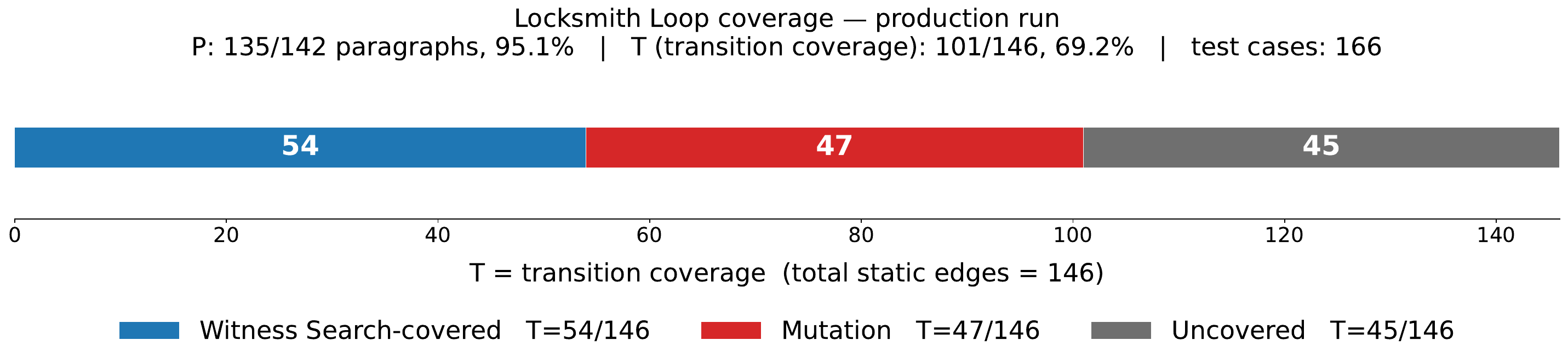}
\caption{Locksmith Loop coverage progression on the
production-shape run, rendered analogously to
Figure~\ref{fig:progression}. Of 146 static paragraph connections,
the baseline Witness Search phase covers 54 (blue); the Mutation phase
adds 47 more (red), bringing the total to
$\mathrm{T}{=}101/146$ ($69.2\%$). Paragraph coverage reaches
$\mathrm{P}{=}135/142$ ($95.1\%$) across 166 test cases.}
\label{fig:prod-progression}
\end{figure*}

\begin{figure}[t]
\centering
\includegraphics[width=\linewidth]{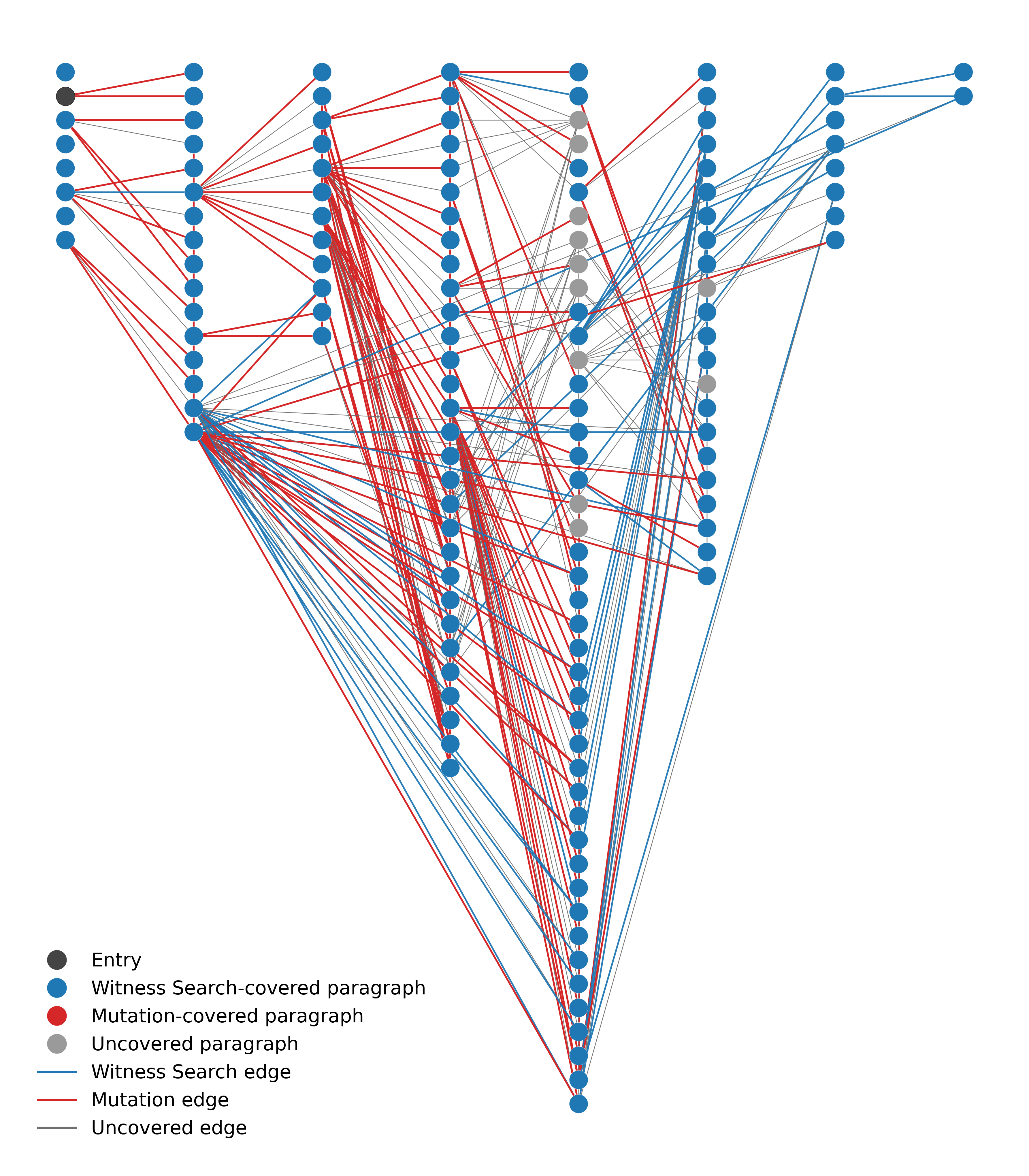}
\caption{Control-flow graph of the production program. Same node and edge color conventions as Figure~\ref{fig:dag}. To reduce clutter, we omit COBOL fall-through semantics; therefore, not all execution paths are represented in the DAG.}
\label{fig:prod-dag}
\end{figure}

The production run plateaued short of 100\% mainly due to structural and external-coupling complexities that may be addressed via manual (human) intervention. Upon inspecting the still-uncovered paragraphs, we consistently identified a pattern: remaining paths sat behind multi-step paragraph chains where control variables are set incrementally across earlier sections. For the probes to enter the remaining paragraphs, the tests must make the fake files and database return an exact coordinated sequence of results, and flip internal switches early. We will continue to build on our research to further analyze and solve such scenarios. 
Nevertheless, we have not yet encountered any blockers, and we foresee, at today's pace of AI improvement, reaching complete paragraph and transition coverage. Even without full coverage, the method brings significant improvement to code pre-integration testing. The aim is to bring migrated Java code close to parity with the legacy COBOL code early in the Software Development Lifecycle (SDLC).


%% file: sections/threats_intro.tex
\label{sec:threats}

We discuss five threats to validity, in order of severity from the perspective of generalizing these claims to
other COBOL migrations.

%% file: sections/threats_external.tex
\paragraph{External validity: case-study scope}
The experiment is conducted on three programs: two open-source batch COBOL programs and one production-shaped run.

The detailed case allows us to demonstrate that the Mutation phase reaches paragraphs and transitions that input-space search alone cannot: $8$ of the $24$ covered paragraphs and $19$ of the $37$ covered transitions. This should be treated as evidence from these cases, not as a general claim about all COBOL systems. To make this claim stronger, we would need to replicate the experiment across additional program families, including IMS, database-heavy SQL, and CICS systems.


%% file: sections/threats_construct.tex
\paragraph{Construct validity: branch coverage as a proxy}

We use branch coverage of the instrumented mock as the primary
quantitative measure. Branch coverage is a well-known but imperfect
proxy for fault-detection effectiveness: a high-coverage test suite
need not detect all classes of fault, and a small set of well-chosen
mutation-tested cases can find faults a higher-coverage suite misses,
a tension long documented in the fuzzing-and-coverage
literature~\cite{ding21ossfuzz}.
We mitigate this by anchoring acceptance on cross-language parity
rather than coverage alone: a test case that adds a branch but
diverges between COBOL and Java is not accepted,
but we do not claim that the resulting suite is sufficient on its
own for migration sign-off. It is a quantitatively comparable
yardstick against the bare-baseline starting point, no more.


%% file: sections/threats_internal.tex
\paragraph{Internal validity: parity preserves bugs}
Anchoring on the legacy COBOL as the \emph{operational reference}
means the loop validates whether the migrated Java reproduces the
legacy implementation's behavior, including its bugs. In a 1:1
migration context this is the desired property, because maintainers will
own the Java code post-migration and any behavioral drift would
surface as a regression they would have to investigate, but it
is genuinely a different claim from ``the program is correct.'' The
operational reference is the truth of \emph{compatibility}, not the
truth of \emph{semantics}. If the eventual goal is correctness
verification rather than compatibility migration, the Parity Gate
alone is insufficient and an external specification oracle must be
introduced.


%% file: sections/threats_overfitting.tex
\paragraph{Overfitting to harness structure}
The Locksmith Loop can only learn from the source programs it has seen, 
so it may not capture structures or execution patterns that
are absent from the initial sample. As a result, programs from application
families not represented in the input sample may require additional 
work to model their structure accurately. Because the current testing 
was scoped to a small sample of programs, it may need to be extended 
to support broader program families.

%% file: sections/threats_java_oracle.tex
\paragraph{Generated Java quality and oracle completeness}
The Parity Gate is only as strong as its equivalence axes
(Section~\ref{sec:parity}). We use three: paragraph-entry set,
ordered stub-operation log, and terminal observable state. These
catch the classes of divergence we have encountered in practice
(missing paragraph traversal, early termination, mis-set variables,
wrong dispatch order), but the gate is silent on, for example,
intermediate state during execution between observable checkpoints,
floating-point rounding modes that differ between COBOL and Java
runtimes, and exception types raised by JVM-only error paths. A
generated Java target could in principle pass our gate while exhibiting
behavioral drift on an axis we don't check. We have not yet
encountered such a case but cannot rule it out from a finite test
campaign, and a production deployment would require the gate's
equivalence axes to be reviewed against the migration's specific
risk profile.


%% file: sections/future_work.tex
\label{sec:future}

We present three  opportunities across expansion, generalization, and evaluation.

\noindent\textbf{1. Expand the number of COBOL-Java transformations}

The current evidence is drawn from two open-source COBOL programs and one
internal production-like COBOL program. Future work to evaluate the methodology
across additional COBOL-to-Java program pairs will reveal whether 
agent-discovered techniques generalise across applications.

\noindent\textbf{2. Generalization of the Locksmith pattern into other domains}

Applying the Locksmith pattern to other pairwise transformation objectives would
greatly expand the potential range use cases it could be support.
Candidate systems would need to satisfy two requirements: first, they must expose
a mechanism to detect system behavior such as logs, traces or system output; second, they must
have a migrator that can be modified to preserve behavioral parity during observation. 

This is consistent with work on automated COBOL-to-Java validation, which emphasizes that 
translated COBOL cannot be trusted without checking semantic equivalence against the original 
source \cite{kumar25cobol2java}.

Future work should explore feasibility of applying this pattern in other domains, such as C\#-to-Java
or transformations involving database query and definition languages such as SQL.

\noindent\textbf{3. Evaluation and improvement of agentic intervention}

Proof of improving agentic intervention is an area that has high potential upside
on the impact of improving branch coverage and reducing token cost. Future work could cover
creation of golden datasets, checkpoints and set up of controlled evaluation parameters to 
measure outcomes such as token cost, task success rate, branch coverage and tool selection accuracy 
under realistic conditions. 

Experiments could focus on different areas of the technology including optimal skill synthesis, 
minimization of necesarry mutations, improved search strategy and improved tool selection.

Future work should also explore whether mutations discovered by agents can be 
extracted and integrated into mutating search-space algorithms such as T-Fuzz~\cite{peng18tfuzz} and work on 
LLM-synthesized mutators ~\cite{wang25mut4all}.


%% file: sections/conclusion.tex
\label{sec:conclusion}

The Locksmith Loop alternates a six-algorithm Witness Search sweep
with parity-preserving mutation, extending search into execution regions
that input-space algorithms alone cannot reach. After each successful
mutation, the loop restarts Witness Search and recursively explores the
newly reachable region. This expanded reach enables higher coverage during parity testing
between COBOL sources and their generated Java targets.

The remaining coverage gaps appear to reflect the current limits of the
Authoring Layer rather than a structural limitation of the Locksmith Loop.
They arose from occasional mutation-authoring errors and failures to transfer
learned strategies to related scenarios and were resolved with human
intervention. The deterministic core of the
Locksmith Loop was designed to operate effectively even when the
agentic Authoring Layer is supplemented by human assistance.
Section~\ref{sec:future} outlines how we plan to extend the
evaluation to more program pairs, generalize the pattern to other
migrations, and strengthen the agentic intervention. In practice, 
the Locksmith Loop has substantially improved our ability
to identify behavioral-drift bugs in migrated code before integration
testing. To our knowledge, the coverage achieved here exceeds that of
any prior method for the same migration setting. We expect stronger
frontier models to increase coverage further and reduce the manual
effort required for trustworthy migration.


%% file: references.bib
@article{kuhn04combinatorial,
  author = {D. Richard Kuhn and Dolores R. Wallace and Albert M. Gallo},
  title = {Software Fault Interactions and Implications for Software Testing},
  journal = {IEEE Transactions on Software Engineering},
  volume = {30},
  number = {6},
  pages = {418--421},
  year = {2004},
  doi = {10.1109/TSE.2004.24}
}

@article{chen10art,
  author = {Tsong Yueh Chen and Fei-Ching Kuo and Robert G. Merkel and T. H. Tse},
  title = {Adaptive Random Testing: The {ART} of Test Case Diversity},
  journal = {Journal of Systems and Software},
  volume = {83},
  number = {1},
  pages = {60--66},
  year = {2010},
  doi = {10.1016/j.jss.2009.02.022}
}

@article{auer02ucb1,
  author = {Peter Auer and Nicol{\`o} Cesa-Bianchi and Paul Fischer},
  title = {Finite-time Analysis of the Multiarmed Bandit Problem},
  journal = {Machine Learning},
  volume = {47},
  number = {2--3},
  pages = {235--256},
  year = {2002},
  doi = {10.1023/A:1013689704352}
}

@misc{mouret15mapelites,
  author = {Jean-Baptiste Mouret and Jeff Clune},
  title = {Illuminating Search Spaces by Mapping Elites},
  year = {2015},
  note = {arXiv:1504.04909},
  url = {https://arxiv.org/abs/1504.04909}
}

@inproceedings{sfikas21mcelites,
  author = {Konstantinos Sfikas and Antonios Liapis and Georgios N. Yannakakis},
  title = {Monte Carlo Elites: Quality-Diversity Selection as a Multi-Armed Bandit Problem},
  booktitle = {Proceedings of the Genetic and Evolutionary Computation Conference},
  series = {GECCO '21},
  pages = {180--188},
  publisher = {ACM},
  year = {2021},
  doi = {10.1145/3449639.3459321}
}

@article{arcuri19mio,
  author = {Andrea Arcuri},
  title = {Test Suite Generation with the Many Independent Objective ({MIO}) Algorithm},
  journal = {Information and Software Technology},
  volume = {104},
  pages = {195--206},
  year = {2018},
  doi = {10.1016/j.infsof.2018.05.003}
}

@article{manes18fuzzing,
  author = {Valentin J. M. Man{\`e}s and HyungSeok Han and Choongwoo Han and Sang Kil Cha and Manuel Egele and Edward J. Schwartz and Maverick Woo},
  title = {The Art, Science, and Engineering of Fuzzing: A Survey},
  journal = {IEEE Transactions on Software Engineering},
  volume = {47},
  number = {11},
  pages = {2312--2331},
  year = {2021},
  doi = {10.1109/TSE.2019.2946563}
}

@inproceedings{peng18tfuzz,
  author = {Hui Peng and Yan Shoshitaishvili and Mathias Payer},
  title = {{T-Fuzz}: Fuzzing by Program Transformation},
  booktitle = {2018 IEEE Symposium on Security and Privacy (SP)},
  pages = {697--710},
  publisher = {IEEE},
  year = {2018},
  doi = {10.1109/SP.2018.00056}
}

@inproceedings{ding21ossfuzz,
  author = {Zhen Yu Ding and Claire Le Goues},
  title = {An Empirical Study of {OSS-Fuzz} Bugs},
  booktitle = {2021 IEEE/ACM 18th International Conference on Mining Software Repositories (MSR)},
  pages = {131--142},
  publisher = {IEEE},
  year = {2021},
  doi = {10.1109/MSR52588.2021.00026}
}

@article{baldoni16symexec,
  author = {Roberto Baldoni and Emilio Coppa and Daniele Cono D'Elia and Camil Demetrescu and Irene Finocchi},
  title = {A Survey of Symbolic Execution Techniques},
  journal = {ACM Computing Surveys},
  volume = {51},
  number = {3},
  pages = {50:1--50:39},
  year = {2018},
  doi = {10.1145/3182657}
}

@article{danglot17amplification,
  author = {Benjamin Danglot and Oscar Luis Vera-P{\'e}rez and Zhongxing Yu and Andy Zaidman and Martin Monperrus and Benoit Baudry},
  title = {A Snowballing Literature Study on Test Amplification},
  journal = {Journal of Systems and Software},
  volume = {157},
  pages = {110398},
  year = {2019},
  doi = {10.1016/j.jss.2019.110398}
}

@article{danglot18dspot,
  author = {Benjamin Danglot and Oscar Luis Vera-P{\'e}rez and Benoit Baudry and Martin Monperrus},
  title = {Automatic Test Improvement with {DSpot}: A Study with Ten Mature Open-Source Projects},
  journal = {Empirical Software Engineering},
  volume = {24},
  pages = {2603--2635},
  year = {2019},
  doi = {10.1007/s10664-019-09692-y}
}

@inproceedings{fraser11evosuite,
  author = {Gordon Fraser and Andrea Arcuri},
  title = {{EvoSuite}: Automatic Test Suite Generation for Object-Oriented Software},
  booktitle = {Proceedings of the 19th ACM SIGSOFT Symposium and the 13th European Conference on Foundations of Software Engineering},
  series = {ESEC/FSE '11},
  pages = {416--419},
  publisher = {ACM},
  year = {2011},
  doi = {10.1145/2025113.2025179}
}

@inproceedings{roslan22evosuiteamp,
  author = {Muhammad Firhard Roslan and Jos{\'e} Miguel Rojas and Phil McMinn},
  title = {An Empirical Comparison of {EvoSuite} and {DSpot} for Improving Developer-Written Test Suites with Respect to Mutation Score},
  booktitle = {Search-Based Software Engineering},
  series = {Lecture Notes in Computer Science},
  volume = {13711},
  pages = {19--34},
  publisher = {Springer},
  year = {2022},
  doi = {10.1007/978-3-031-21251-2_2}
}

@article{harman04testability,
  author = {Mark Harman and Lin Hu and Robert M. Hierons and Joachim Wegener and Harmen Sthamer and Andr{\'e} Baresel and Marc Roper},
  title = {Testability Transformation},
  journal = {IEEE Transactions on Software Engineering},
  volume = {30},
  number = {1},
  pages = {3--16},
  year = {2004},
  doi = {10.1109/TSE.2004.1265732}
}

@article{garousi18testability,
  author = {Vahid Garousi and Michael Felderer and Feyza Nur K{\i}l{\i}{\c c}aslan},
  title = {A Survey on Software Testability},
  journal = {Information and Software Technology},
  volume = {108},
  pages = {35--64},
  year = {2019},
  doi = {10.1016/j.infsof.2018.12.003}
}

@inproceedings{stephens16driller,
  author = {Nick Stephens and John Grosen and Christopher Salls and Andrew Dutcher and Ruoyu Wang and Jacopo Corbetta and Yan Shoshitaishvili and Christopher Kruegel and Giovanni Vigna},
  title = {Driller: Augmenting Fuzzing Through Selective Symbolic Execution},
  booktitle = {Proceedings of the 23rd Annual Network and Distributed System Security Symposium},
  series = {NDSS 2016},
  publisher = {Internet Society},
  year = {2016},
  doi = {10.14722/NDSS.2016.23368}
}

@article{chen18metamorphic,
  author = {Tsong Yueh Chen and Fei-Ching Kuo and Huai Liu and Pak-Lok Poon and Dave Towey and T. H. Tse and Zhi Quan Zhou},
  title = {Metamorphic Testing: A Review of Challenges and Opportunities},
  journal = {ACM Computing Surveys},
  volume = {51},
  number = {1},
  pages = {4:1--4:27},
  year = {2018},
  doi = {10.1145/3143561}
}

@article{etemadi24mokav,
  author = {Khashayar Etemadi and Bardia Mohammadi and Zhendong Su and Martin Monperrus},
  title = {Mokav: Execution-Driven Differential Testing with {LLMs}},
  journal = {Journal of Systems and Software},
  volume = {230},
  pages = {112571},
  year = {2025},
  doi = {10.1016/j.jss.2025.112571}
}

@misc{rao24diffspec,
  author = {Nikitha Rao and Elizabeth Gilbert and Tahina Ramananandro and Nikhil Swamy and Claire Le Goues and Sarah Fakhoury},
  title = {{DiffSpec}: Differential Testing with {LLMs} Using Natural Language Specifications and Code Artifacts},
  year = {2024},
  note = {arXiv:2410.04249},
  url = {https://arxiv.org/abs/2410.04249}
}

@article{zhang23apr,
  author = {Quanjun Zhang and Chunrong Fang and Yuxiang Ma and Weisong Sun and Zhenyu Chen},
  title = {A Survey of Learning-Based Automated Program Repair},
  journal = {ACM Transactions on Software Engineering and Methodology},
  volume = {33},
  number = {2},
  pages = {55:1--55:69},
  year = {2024},
  doi = {10.1145/3631974}
}

@article{yang25llmapr,
  author = {Boyang Yang and Zijian Cai and Feng Liu and Bach Le and Lingming Zhang and T{\'e}gawend{\'e} F. Bissyand{\'e} and Yang Liu and Haoye Tian},
  title = {A Survey of {LLM}-Based Automated Program Repair: Taxonomies, Design Paradigms, and Applications},
  journal = {ArXiv},
  volume = {abs/2506.23749},
  year = {2025},
  url = {https://api.semanticscholar.org/CorpusID:280010745}
}

@inproceedings{bouzenia24repairagent,
  author = {Islem Bouzenia and Premkumar Devanbu and Michael Pradel},
  title = {{RepairAgent}: An Autonomous, {LLM}-Based Agent for Program Repair},
  booktitle = {2025 IEEE/ACM 47th International Conference on Software Engineering (ICSE)},
  pages = {2188--2200},
  publisher = {IEEE},
  year = {2025},
  doi = {10.1109/ICSE55347.2025.00157}
}

@inproceedings{jimenez23swebench,
  author = {Carlos E. Jimenez and John Yang and Alexander Wettig and Shunyu Yao and Kexin Pei and Ofir Press and Karthik Narasimhan},
  title = {{SWE}-bench: Can Language Models Resolve Real-World {GitHub} Issues?},
  booktitle = {The Twelfth International Conference on Learning Representations},
  year = {2024},
  url = {https://openreview.net/forum?id=VTF8yNQM66}
}

@inproceedings{yang24sweagent,
  author = {John Yang and Carlos E. Jimenez and Alexander Wettig and Kilian Lieret and Shunyu Yao and Karthik Narasimhan and Ofir Press},
  title = {{SWE}-agent: Agent-Computer Interfaces Enable Automated Software Engineering},
  booktitle = {Advances in Neural Information Processing Systems 37},
  pages = {50528--50652},
  year = {2024},
  doi = {10.52202/079017-1601}
}

@article{schafer23testpilot,
  author = {Max Sch{\"a}fer and Sarah Nadi and Aryaz Eghbali and Frank Tip},
  title = {An Empirical Evaluation of Using Large Language Models for Automated Unit Test Generation},
  journal = {IEEE Transactions on Software Engineering},
  volume = {50},
  number = {1},
  pages = {85--105},
  year = {2024},
  doi = {10.1109/TSE.2023.3334955}
}

@inproceedings{alshahwan24testgenllm,
  author = {Nadia Alshahwan and Jubin Chheda and Anastasia Finegenova and Beliz Gokkaya and Mark Harman and Inna Harper and Alexandru Marginean and Shubho Sengupta and Eddy Wang},
  title = {Automated Unit Test Improvement Using Large Language Models at {Meta}},
  booktitle = {FSE 2024: Companion Proceedings of the 32nd ACM International Conference on the Foundations of Software Engineering},
  pages = {185--196},
  publisher = {ACM},
  year = {2024},
  doi = {10.1145/3663529.3663839}
}

@misc{wang24llmmutation,
  author = {Bo Wang and Mingda Chen and Youfang Lin and Mike Papadakis and Jie M. Zhang},
  title = {An Exploratory Study on Using Large Language Models for Mutation Testing},
  year = {2024},
  note = {arXiv:2406.09843},
  url = {https://arxiv.org/abs/2406.09843}
}

@misc{wang25mut4all,
  author = {Bo Wang and Pengyang Wang and Chong Chen and Qi Sun and Jieke Shi and Chengran Yang and Ming Deng and Youfang Lin and Zhou Yang and Junjie Chen and Jun Sun and David Lo},
  title = {{Mut4All}: Fuzzing Compilers via {LLM}-Synthesized Mutators Learned from Bug Reports},
  year = {2025},
  note = {arXiv:2507.19275v2},
  url = {https://arxiv.org/abs/2507.19275}
}

@inproceedings{kumar25cobol2java,
  author = {Atul Kumar and Diptikalyan Saha and Toshikai Yasue and Kohichi Ono and Saravanan Krishnan and Sandeep Hans and Fumiko Satoh and Gerald Mitchell and Sachin Kumar},
  title = {Automated Validation of {COBOL} to {Java} Transformation},
  booktitle = {Proceedings of the 39th IEEE/ACM International Conference on Automated Software Engineering},
  series = {ASE '24},
  pages = {2415--2418},
  publisher = {ACM},
  year = {2024},
  doi = {10.1145/3691620.3695365}
}

@inproceedings{gandhi24coboljava,
  author = {Shubham Gandhi and Manasi Patwardhan and Jyotsana Khatri and Lovekesh Vig and Raveendra Kumar Medicherla},
  title = {Translation of Low-Resource {COBOL} to Logically Correct and Readable {Java} Leveraging High-Resource {Java} Refinement},
  booktitle = {Proceedings of the 1st International Workshop on Large Language Models for Code},
  series = {LLM4Code '24},
  pages = {46--53},
  publisher = {ACM},
  year = {2024},
  doi = {10.1145/3643795.3648388}
}

@inproceedings{hans25cobol2javatesting,
  author = {Sandeep Hans and Atul Kumar and Toshikai Yasue and Kouichi Ono and Saravanan Krishnan and Devika Sondhi and Fumiko Satoh and Gerald Mitchell and Sachin Kumar and Diptikalyan Saha},
  title = {Automated Testing of {COBOL} to {Java} Transformation},
  booktitle = {FSE 2025: Companion Proceedings of the 33rd ACM International Conference on the Foundations of Software Engineering},
  pages = {227--237},
  publisher = {ACM},
  year = {2025},
  doi = {10.1145/3696630.3728548}
}

@inproceedings{chakravarthy26enterprise,
  author = {Venkatesan Chakaravarthy and Anamitra Roy Choudhury and Dinesh Garg and Vini Kanvar and Shivmaran Pandian and Aditya Raghuvanshi and Yogish Sabharwal and Amith Singhee},
  title = {Enterprise-Scale {COBOL}-to-{Java} Translation: {LLMs} Augmented with Program Analysis},
  booktitle = {Proceedings of the 48th IEEE/ACM International Conference on Software Engineering: Software Engineering in Practice},
  series = {ICSE-SEIP '26},
  pages = {775--785},
  year = {2026},
  doi = {10.1145/3786583.3786915}
}

@misc{dau24xmainframe,
  author = {Anh T. V. Dau and Hieu Trung Dao and Anh Tuan Nguyen and Hieu Trung Tran and Phong X. Nguyen and Nghi D. Q. Bui},
  title = {{XMainframe}: A Large Language Model for Mainframe Modernization},
  year = {2024},
  note = {arXiv:2408.04660},
  url = {https://arxiv.org/abs/2408.04660}
}

@misc{dau26cobolcoder,
  author = {Anh T. V. Dau and Shin Hwei Tan and Jinqiu Yang and Nghi D. Q. Bui and Anh Tuan Nguyen},
  title = {{COBOL}-Coder: Domain-Adapted Large Language Models for {COBOL} Code Generation and Translation},
  year = {2026},
  note = {arXiv:2604.03986},
  url = {https://arxiv.org/abs/2604.03986}
}

@misc{carddemo,
  author = {{Amazon Web Services}},
  title = {CardDemo - Mainframe Credit Card Management Application},
  howpublished = {GitHub repository},
  year = {2022},
  url = {https://github.com/aws-samples/aws-mainframe-modernization-carddemo}
}
